\documentclass{aa}
\usepackage{graphics}
\usepackage{amsfonts}
\include{epsf}
\include{epstopdf}
\newcommand{\bfo}[1]{\mbox{\boldmath $#1$}}
\def\bvarphi{\mbox{\boldmath $\varphi$}}

\begin{document}
\newcommand{\beq}{\begin{equation}}
\newcommand{\eeq}{\end{equation}}
%****** Begin Definitions *************************
%\def {\sffamily}
\def\la{\hbox{\raise.35ex\rlap{$<$}\lower.6ex\hbox{$\sim$}\ }}
\def\ga{\hbox{\raise.35ex\rlap{$>$}\lower.6ex\hbox{$\sim$}\ }}
\def\runit{\hat {\bf  r}}
\def\phunit{\hat {\bfo \bvarphi}}
\def\etaunit{\hat {\bfo \eta}}
\def\zunit{\hat {\bf z}}
\def\zetaunit{\hat {\bfo \zeta}}
\def\xiunit{\hat {\bfo \xi}}
\def\beq{\begin{equation}}
\def\eeq{\end{equation}}
\def\beqa{\begin{eqnarray}}
\def\eeqa{\end{eqnarray}}
\def\sub#1{_{_{#1}}}
\def\order#1{{\cal O}\left({#1}\right)}
\newcommand{\sfrac}[2]{\small \mbox{$\frac{#1}{#2}$}}
%****** End Definitions ***************************
%
%
\title{{A shallow-water theory for annular sections of Keplerian Disks}}

\author{O.M. Umurhan\inst{1,2,3}}

   \offprints{O.M. Umurhan \email{mumurhan@ccsf.edu}}

   \institute{
   Astronomy Unit, School of Mathematical Sciences, Queen Mary
   University of London, London E1 4NS, U.K.\
     \and
     Department of Geophysics and Space Sciences, Tel-Aviv University,
     Tel-Aviv, Israel\
     \and
        Astronomy Department, City College of San Francisco,
      San Francisco, CA 94112, USA\
}

\date{Received 3 February 2008 ; accepted 21 July 2008}

% \abstract{}{}{}{}{}
% 5 {} token are mandatory

  \abstract
  {We present a scaling argument that we develop into a shallow water theory of non-axisymmetric
  disturbances in annular sections of thin Keplerian disks.}
  {We develop a theoretical approach to understand physically
  the relationship between two-dimensional vortex dynamics that is known and
  their three-dimensional counterparts in Keplerian disks.}
  {Using asymptotic scaling arguments, varicose disturbances
  of a Keplerian disk are considered on radial and vertical scales
  consistent with the height of the disk while the
  azimuthal scales are the full $2\pi$ angular extent of the disk.  For simplicity
  perturbations are assumed to be homentropic according to
  a polytropic equation of state.  The
  timescales considered are long compared to the local
  disk rotation time.}
  { The scalings relate to dynamics
  that are radially geostrophic and vertically hydrostatic.
  A potential vorticity  quantity emerges
  and is shown to be conserved in a Lagrangian sense.  Uniform
  potential vorticity linear solutions
  are explored and the theory
  is shown to contain an incarnation of the strato-rotational
  instability under channel flow conditions.  {Linearized solutions
  of a single defect on an infinite domain are developed and shown to support
  a propagating Rossby edgewave.}
  Linear non-uniform potential vorticity solutions are
  also developed and shown to be similar in
  some respects to the dynamics of strictly two-dimensional
  inviscid flows.
  The relationship
  of the scalings and some of the resulting dynamics
  are considered with respect to other approximations employed
  in the literature.   {Based on the framework of
  this theory, arguments based on geophysical notions
  are presented to support the
  assertion that the strato-rotational instability is in
  a generic class of barotropic/baroclinic potential vorticity instabilities.}
  Extensions of this formalism are also proposed.}
  {The shallow water formulation achieved by the asymptotic theory developed
  here opens a new
  approach in studying disk dynamics.}
  % context heading (optional)
  % {} leave it empty if necessary
  %   {Nonsense}
  %   {Nonsense}
  %   {Nonsense}
  %   { Nonsense}
  %  {Nonsense}

\titlerunning{Disk shallow water theory}

\keywords{accretion, accretion disks -- instabilities}

  \maketitle
%________________________________________________________________________

\section{Introduction}
The question of the formation of coherent structures
in Keplerian disks has received increased
attention in the past few years.  Their
importance has been recognized in playing a role
in a number of processes including angular
momentum transport and the accumulation of dust
in the process of planet building.  These coherent
structures are persistent vortices and
understanding their generation, structure, dynamics and
lifetimes has been the focus of
a number of studies both in 2D
(e.g. Godon \& Livio, 1999, Bracco et al., 1999,
Umurhan \& Regev, 2004, Johnson \& Gammie, 2005,
Petersen et al. A, 2007, Petersen et al. B, 2007, Bodo et al. 2007, Lithwick, 2007 A)
and recent 3D studies (Barranco \& Marcus, 2005, Lithwick, 2007 B).
\par
There are many facets to the nature of vortices in strongly
sheared and rotating environments like astrophysical disks.
One thing that most two-dimensional
studies have shown, and which are in qualitative agreement with each
other, is that from a patch of noisy initial conditions anti-cyclonic vortices
naturally emerge and these structures are usually the sole occupants of
radially demarcated zones.  This generic phenomenon is
observed in 2D shearing box simulations (Umurhan \& Regev, 2004, Johnson \& Gammie, 2005
Lithwick, 2007 A)
and 2D cylindrical studies as well (Godon \& Livio, 1999,
Bracco et al., 1999,
Petersen et al. A-B, 2007, Bodo et al. 2007).
In all of these examples a vorticity (or potential vorticity) quantity
primarily characterizes the flow dynamics.  This vorticity may evolve (nearly)
conservatively in a Lagrangian sense
(e.g. Godon \& Livio, 1999, Bracco et al. 2000, Umurhan \& Regev, 2004, Lithwick, 2007 A) or may have a source term due
to baroclinicity leading to, for example, Rossby wave-like dynamics (i.e. wave
generation due to radial gradients in temperature or
entropy as in Lovelace et al., 1999, Li et al. 2000, Klahr \&
Bodenheimer, 2003, Petersen et al. A-B, 2007),
or other wave generation processes of some type (see e.g.
the compressible studies of Johnson \& Gammie 2005, Bodo et al., 2007).
\par
The dynamics of vortices in 3D simulations is less clear and
the picture of what is occurring under these circumstances
is only now emerging.  Lithwick (2007 B) has demonstrated
that columnar anticyclonic vortices can manifest themselves in
3D shearing box simulations with no external gravity.  It was also shown in
that study that the survival of the vortex column depends
on the height of the column itself - if it is too large
secondary instabilities may develop causing a toppling over
of the structure, otherwise, if it is short enough
the structure continues to persist.
 In an anelastic stratified fluid model
evolved in a shearing box
with the presence of
an external disk-vertical component of gravity, Barranco \& Marcus (2005) show that
columnar anticyclonic vortices show a tendency to breakdown
but are replaced with anticyclonic structures of restricted vertical extent
located away from the disk midplane.  The way that this
process emerges is subtle and a full mechanical
understanding of it still awaits - perhaps aided by a shallow water model
description.\par
Shallow water models are effective at
capturing salient physical effects involving
stratification without having to resort to full 3D
analyses and simulations. In this way they have proven to be
indispensible tools in furthering our physical understanding of
oceanic and atmospheric
flows (Pedlosky, 1987) and have recently been used
in studies of exoplanet atmospheres (Langton \& Laughlin, 2007, Cho et al., 2007).

\par
In a series of studies (Balmforth et al. 1992, Balmforth et al., 1993,
Balmforth et al. ,1995, Balmforth \& Spiegel, 1996)
a shallow water formalism was presented and analyzed to
study the axisymmetric structure and dynamics of thin cold disks
both with and without the effects of self-gravity.  This
formalism proved to be especially effective at capturing and
describing the dynamical response of self-gravitating disk systems.
In a limit akin to a shallow water reduction, the linearized equations of motion
of small sections of accretion tori
have been shown to lead to interesting non-axisymmetric dynamics including
processes such as the Papaloizou-Pringle
instability ({\em PPI}, Papalouizou \& Pringle, 1984,1985, Goldreich et al., 1986).
We shall return to this in the Discussion section. \par

Owing to the ubiquity of (potential) vortex dynamics in 2D
and the appearance of long-lived coherent structures in
3D simulations it is worthwhile to develop a
framework to better understand the formation
of vortices in 3D scenarios by comparison to
the processes leading to 2D vortex structures.
A shallow water theory appropriate for
non-axisymmetric disturbances of an accretion disk should serve
a critical intermediary role in understanding the physical
development of such processes.
In this context, meteorological studies ought not to be overlooked:
potential vorticity dynamics ( {sometimes referred to in this
  work as ``PV"}) and its role
in the generation and development of planetary storms
in the upper atmosphere have been long appreciated
(Hoskins et al., 1985, Bishop \& Davies, 1994, and many others).
\par
The road to such a shallow water formalism is guided in part by
the results thus far established in the literature.
Especially important is the tendency seen in
2D cylindrical
studies for (anticyclonic) vortices to either fill out or be the sole occupant of the azimuthal
scale of their radially restricted domain (unambiguous
examples of this may be found in Godon \& Livio, 1999, Petersen et al. A, 2007, and
Petersen et al. B, 2007).
When viewed in a
reference frame moving with a steady vortex the scale of the
azimuthal velocities (``$\tilde{{\cal U}}$") are larger than the corresponding
radial velocities in proportion to the ratio of
the azimuthal to radial length scales of the vortex structure.
Assuming that the azimuthal scale is the radial position
where the vortex is found (``$R_0$") then
the circulation time of these vortex structures is $\sim R_0/\tilde{{\cal U}}$.
Such vortices in thin (cold) disks, whose azimuthal velocity
scales are no greater than the typical sound speeds,
will have time scales which
are $1/\epsilon$ longer than the typical rotation time of the disk
at the location of the vortex, where $\epsilon$ is the ratio
of the local soundspeed to the local Keplerian velocity.  In typical
cold disks this number is on the order of $1/20$.
\par

We therefore reanalyze
the full equations of motion (in cylindrical coordinates)
by considering the dynamics taking place in
an annular section of the disk instead of the usual
box section known as the Shearing Box (Goldreich \& Lynden-Bell,1965).
The radial and vertical scales are scaled by the disk height,
$H_0 = \epsilon R_0$, where (as above) $R_0$ is the radial scale of the disk.
The azimuthal length scale considered is the full $2\pi R_0$ complement of the disk.
The velocity perturbations sitting atop the basic
flow (which we assume to be Keplerian, i.e. rotationally supported)
are such that the azimuthal velocity fluctuations scale with the local
soundspeed while the radial and vertical velocities are much smaller than
this by order $\epsilon$.  Most importantly, temporal variations
do not scale by the usual rotation time scale (given by the inverse
of the Keplerian rotation frequency $\bar\Omega_0$ of the disk
at $R_0$) but
on a time scale which is $1/\epsilon$ longer.  In order to
affect a shallow water reduction in a transparent way
without entirely dispensing with the compressible nature of the gas itself
we assume here that the disk gas behaves homentropically where,
in particular, we assume a polytropic equation of state characterized
by the index $\gamma$ (as in Salby 1989, Cho et al. 2007 and see also
the earlier studies of Papalouizou \& Pringle, 1984, 1985, Goldreich et al. 1986).
\par
These adopted scalings, especially the slow timescale, lead to conditions in which disturbances
are vertically hydrostatic.  It also follows that disturbances are geostrophic
but only in the radial direction.  This means to say that the dynamics
are everywhere in radial geostrophic balance, i.e.
radial pressure gradients balancing Coriolis effects.
These circumstances are similar
to the situation encountered resulting from similar scalings of the
standard equations of the shearing box (e.g. the Large-Shearing
Box limit in Umurhan \& Regev, 2004, Umurhan, 2006).  This type
of leading order radial geostrophic balance has been
encountered in prior related work as well (Balmforth \& Spiegel, 1996,
Barranco et al., 2000).
\par
The main result of this analysis then shows that in this
asymptotic limit there exists a potential vorticity
quantity which is conserved in a Lagrangian sense in
the radial and azimuthal directions.  In other words
we have
\beq
\frac{D\Xi}{Dt} = 0,\qquad
\Xi = \frac{\Omega_0(2-q) + \partial_x v}{h^{\frac{\gamma+1}{\gamma-1}}},
\label{DSWE_1}
\eeq
where $\Xi$ is the potential vorticity.
The azimuthal velocity $v$
represents perturbations atop the basic
(local) Keplerian flow
%($= -q\Omega_0 x$)
and where $h(x,y,t)$ is the height of the disk measured
from the disk midplane.  The Lagrangian time derivative operation is restricted
to the two horizontal directions:
\[
\frac{D}{Dt} \equiv \partial_t + u\partial_x + (v - q\Omega_0 x)\partial_y,
\]
and with $\Omega_0$ measuring the local rotation rate of the disk (here it is scaled to 1).
The basic Keplerian flow is given by $-q\Omega_0 x$ in which $x$ is the local
(Cartesion) radial coordinate and $q=3/2$.  The azimuthal and radial velocities are given
diagnostically in terms of the height of the disk,
\beqa
v &=& {\cal V}(h) \equiv \frac{\Omega_0}{4}\partial_x h^2, \label{DSWE_2} \\
u &=& {\cal P}^{-1}{\cal F}(h), \label{DSWE_3}
\eeqa
in which the operators ${\cal P}$ and ${\cal F}$ are
\beqa
{\cal F}(h) &\equiv&
(\partial_x {\cal V} - q\Omega_0 - 2\Omega_0)\cdot\partial_y Q_h
+2\partial_x\left(\frac{\gamma-1}{\gamma+1}Q_h \partial_y {\cal V}\right), \nonumber \\
 {\cal P} & \equiv & \Omega_0^22(2-q) - 2\Omega_0 {\cal V}\cdot \partial_x
-2\partial_x\left(\frac{\gamma-1}{\gamma+1}Q_h \partial_x \right).
\label{DSWE_4}
\eeqa
with $Q_h \equiv \sfrac{1}{2}\Omega_0^2 h^2$.
\par
The radial and horizontal velocities are {\emph {derived}} to be related to the vertical
linearly varying velocity, (i.e. $w=\Omega_z z$, where $\Omega_z=\Omega_z(x,y,t)$)
via a quasi-anelastic equation of state,
\beq
\partial_x u + \partial_y v + \frac{\gamma+1}{\gamma -1}\partial_z w = 0.
\label{DSWE_5}
\eeq
As such, these equations describe varicose disturbances with respect to the disk
midplane (cf. usage in Balmforth et al., 1992)
since the vertical velocity has odd symmetry with respect
to the midplane.
\par
We use the set (\ref{DSWE_1}-\ref{DSWE_5}) to solve a number of example
problems.
%We first consider the dynamics of uniform
%potential vorticity disturbances restricted to either
%a semi-infinite domain or a channel flow configuration.
%In a linearized study of the semi-infinite domain we find that
%waves are generated due to the inner no-flow boundary condition.\par
%We find that for $\gamma = 3$ one
%may easily write down exact nonlinear solutions and
%we find that the amplitude of these waves obeys a Burgers
%equation showing that in these circumstances shocks
%are expected to form.\par
A linearized study of uniform potential vorticity disturbances in a
channel flow configuration is presented.  Disturbances with
only one solid boundary on a semi-infinite domain manifest as a single
edgewave (Goldreich et al. 1986) propagating along the
wall.  Disturbances on a finite domain sandwiched by
two solid boundaries manifest
as two edgewaves - each associated with their respective
boundary.  These objects are similar to the edgewaves
discussed by Goldreich et al. (1986) rationalizing the Papalouizou-Pringle
instability (Papalouizou \& Pringle, 1985).  These edgewaves
become unstable if the width
of the channel lies between a narrow band of values less
than the height of the
fluid layer.
For these circumstances we
understand this instability as being the incarnation
of the stratorotational instability discussed by
Dubrulle et al. (2004).  \par
 {Linearized disturbances of a radially infinite domain
with a single global potential vorticity defect appear
as a Rossby edgewave propagating along the defect.   When
the potential vorticity on either side of the defect becomes infinite
(negatively) the Rossby edgewave limits to the
edgewave solution associated with the semi-infinite domain
discussed above.  We therefore rationalize the use of
a solid boundary as representing this limiting form.}\par
%We also find exact nonlinear
%solutions for $\gamma = 3$ showing that the amplitude
%of this instability evolves according to two
%coupled Burgers equations although detailed solutions
%are not pursued in this work.\par
We also investigate the dynamics of non-uniform potential
vorticity disturbances subject to the popular
shearing-sheet boundary conditions (Goldreich \& Lynden-Bell, 1965).
We find that the potential vorticity evolves in ways that
are very similar to 2D vorticity disturbances as in
Chagelishvilli et al. (2003) and Umurhan \& Regev (2004).
We find that potential vorticity waves which are leading (in character)
show strong transient growth thanks to the Orr-tilting
mechanism (Schmid \& Henningson, 2000) which is inline with
the 2D studies cited as well as being inline with the
general tendency for this to happen in 3D linearized
disturbances as studied by Korycansky (1992)
and Tevzadze et al. (2004).\par
This work is organized as follows.  In Section 2 we
develop the derivation of (\ref{DSWE_1}-\ref{DSWE_5}).   {In Sections 3-5
we present the linear results of the study of the uniform, single defect and non-uniform
potential vorticity disturbances (respectively). }
In Section 6 we, summarize our results, discuss
the relationship
of the scaling assumptions employed
here in context of other approximations
employed in the literature,
consider some of the dynamical results with
respect to other published results
and we discuss possible future directions and extensions.   {We suggest that despite the use of artificial boundary
conditions in the classical SRI analysis, that the instability could be
possible in real disks supporting regions of oppositely signed
potential vorticity gradients. }

\section{Scaling Analysis and Derivation of the reduced equations}
The full equations of motion
\beqa
\frac{d\rho}{dT} + \rho \nabla\cdot {\bf U} &=& 0, \\
\frac{d {\bf U}}{dT} &=& \frac{1}{\rho}\nabla P - \nabla\Phi, \\
\frac{d S}{dT} &=& 0,
\eeqa
are considered in cylindrical coordinates ($R,\phi,Z$)
where ${\bf U}\equiv (U_R,U_\phi,U_Z)$ and where
the entropy $S \equiv C_V\ln P/\rho^\gamma$ with $\gamma = C_P/C_V$, i.e.
the ratio of specific heats.  The external potential comes from the
central gravitating source and has the form
\[
\Phi \equiv \frac{\bar\Omega_0^2 R_0^2}{\left(\frac{R^2}{R_0^2} + \frac{Z^2}{R_0^2}\right)^{1/2}},
\]
in which the frequency $\bar\Omega_0 = \sqrt{GM/R_0^3}$ is the Keplerian rotation rate at the fiducial
radius $R=R_0$.
The derivation of the equations follows that of Umurhan \& Regev (2004).  We shall
consider a region centered around the radial point $R=R_0$ and $Z=0$.  We transform
our reference frame so that we are rotating with frequency $\bar\Omega_0$ appropriate
for the radial point $R_0$.
We begin by
stating a number of scalings for the various quantities involved.  We assume that at
this point the density of the fluid has a scale $\bar \rho$ and that the pressure
scales as $\bar\rho c_s^2$ where $c_s$ is the sound speed.  Furthermore, we define
the non-dimensional parameter
\[
\epsilon = \frac{c_s}{v_k}, \qquad v_k \equiv \bar\Omega_0 R_0,
\]
where $v_k$ is the Keplerian rotation velocity at $R=R_0$.
We shall assume that $\epsilon$ is less than 1 and ``small".  We shall
consider radial and vertical extent which scale as $\epsilon R_0$ - thus
this is formally a box-like region in these directions.  For the azimuthal
scales we shall assume them to be of the order $R_0$.  In this sense
the domain represents a thin annulus instead of a classic small box like the
shearing box.\par
The departures from the classic shearing box equation derivation (e.g. as in
Umurhan \& Regev, 2004)
appears at this point.
Following the heuristic observations and arguments proffered in the Introduction,
we assume that time scales as $1/\epsilon\bar\Omega_0$
and that the vertical and radial velocities (in the frame moving with frequency
$\bar\Omega_0$) scale as $\epsilon c_s$ while the azimuthal velocity fluctuations scale
as the usual $c_s$.  In summary we assume the following scalings
\beqa
& & R\rightarrow R_0 + \epsilon R_0 x,
\quad Z\rightarrow \epsilon R_0 z, \nonumber \\
& & \phi \rightarrow  y - \bar\Omega_0 t,\quad  T \rightarrow t/(\epsilon\bar\Omega_0),
\nonumber \\
& & U_R \rightarrow \epsilon c_s u, \quad
U_z \rightarrow \epsilon c_s w, \quad
U_\phi \rightarrow  v_k + c_s v_y, \nonumber \\
& & \rho \rightarrow \bar\rho \rho, \quad
P \rightarrow \bar \rho c_s^2 p.
\eeqa
The variables $p,\rho,t,x,z,y,u,w,v_y$ appearing on the righthandsides
of all transformations are understood to be order 1 non-dimensional quantities
corresponding to their dimensional counterparts.  Inserting these expansions
into the governing equations of motion expressed in the rotating frame
yields the following set of equations at lowest order in $\epsilon$,
\beqa
\frac{d\rho}{dt} + \rho\nabla\cdot {\bf u}
 &=& 0 + \order{\epsilon} \label{continuity_eqn}\\
-2\Omega_0 v_y &=& -\frac{1}{\rho}\partial_x p + 2\Omega_0^2 q x + \order{\epsilon^2} \label{x_momentum_eqn}\\
\frac{dv_y}{dt} + 2\Omega_0 u &=& -\frac{1}{\rho}\partial_y p + \order{\epsilon^2}\label{y_momentum_eqn}\\
0&=& -\frac{1}{\rho}\partial_z p - \Omega_0^2 z + \order{\epsilon^2} \label{vert_eqn}\\
\frac{ds}{dt} &=& 0 +  \order{\epsilon}. \label{entropy_eqn}
\eeqa
where $s \equiv \ln p/\rho^\gamma$, $q=3/2$ and $\Omega_0 =1$.  The
(Cartesian) time derivative operation
is given to be
\[
\frac{d}{dt} = \partial_t + u\partial_x + v_y\partial_y + w\partial_z,
\]
while the divergence operation is
\[
\nabla\cdot {\bf u} = \partial_x u + \partial_y v_y + \partial_z w.
\]
We shall assume that the fluid behaves as a barotrope where, in particular,
we assume that it behaves like a polytropic gas, i.e.
\beq
p = K \rho^\gamma,
\eeq
and consequently it means that the fluid layer is homentropic.  This
is to say that the fluid layer is characterized by a constant entropy
given to be $s=\ln K$.  We may now exactly solve for (\ref{vert_eqn})
by defining the fluid enthalpy $Q$ (as in Goldreich et al., 1986),
\beq
dQ = \frac{1}{\rho}dP = K\frac{\gamma}{\gamma-1} d\rho^{\gamma-1}.
\eeq
In terms of the enthalpy (\ref{vert_eqn}) is simply
\beq
\partial_z Q = -\Omega_0^2 z,
\eeq
yielding the solution
\beq
Q = \sfrac{1}{2}\Omega_0^2\left(h^2 - z^2\right).\label{Q_sol}
\eeq
The height $h = h(x,y,t)$ defines where the fluid density and pressure
go to zero and is a function of the horizontal coordinates and time.
\par
The entropy equation (\ref{entropy_eqn}),
the equation of continuity (\ref{continuity_eqn}), the polytropic
gas assumption and the solution to $Q$ (\ref{Q_sol}), all
taken together yield
\beq
\frac{dQ}{dt} + (\gamma-1)Q\nabla\cdot{\bf u} = 0. \label{enthalpy_eqn}
\eeq
For the solutions we shall examine here,
as done in the derivation of shallow water equations,
\emph{we state the ansatz}
that the horizontal velocities have no z-dependencies while
the vertical velocity has a linear z-dependence, i.e.
\beq
u= u(x,y,t), \qquad v_y = v_y(x,y,t),\qquad w = \Omega_z(x,y,t) z.\ \
\eeq
By choosing this form for the vertical velocity we are
restricting attention to varicose disturbances of this
annular disk section.
Explicitly inserting the solution form (\ref{Q_sol}) into
(\ref{enthalpy_eqn}) and making use of the ansatz for $w$ we
find
\beqa
& & \frac{D}{Dt}\left(\sfrac{1}{2}\Omega_0^2 h^2\right) - \Omega_z\Omega_0^2 z^2 \nonumber \\
& & \ \ \ \ (\gamma-1)\sfrac{1}{2}\Omega_0^2(h^2-z^2)
(\partial_x u + \partial_y v_y + \Omega_z) = 0,
\label{midway}
\eeqa
in which
\[
\frac{D}{Dt} \equiv \partial_t + u\partial_x + v_y\partial_y.
\]
On account of their linear independence, we equate to zero the coefficients of
each power of $z$ appearing in (\ref{midway}).  In this particular situation
this yields two equations, the first is the equation for the height $h$
\beq
\frac{Dh}{Dt} = h\Omega_z,  \label{h_eqn}
\eeq
while the second is an expression resembling an anelastic condition
\beq
\partial_x u + \partial_y v_y + \frac{\gamma+1}{\gamma-1} \Omega_z = 0.
\label{anelastic_eqn}
\eeq
We note that the limit $\gamma\rightarrow\infty$ recovers the
condition of incompressible fluid flow.
Addressing (\ref{x_momentum_eqn}) we rewrite the azimuthal velocity
$v_y = -q\Omega_0 x + v$ to find this equation appearing now
as the geostrophic balance discussed in the Introduction,
\beq
-2\Omega_0 v = -\partial_x Q, \label{semi_geostrophy}
\eeq
while (\ref{y_momentum_eqn}) now appears
\beq
\frac{Dv}{Dt} + \Omega_0(2-q)u = -\partial_y Q. \label{v_eqn}
\eeq
Note that the time derivative looks like
\[
\frac{D}{Dt} = (\partial_t - q\Omega_0 x\partial_y) + u\partial_x + v\partial_y.
\]
The set of equations (\ref{h_eqn}-\ref{v_eqn}) form the
basic equations of this asymptotic limit.
\par
We may proceed even further and develop a potential vorticity equation.
We begin by operating on (\ref{semi_geostrophy}) with $\partial_y$ and subtracting
from this the result of operating on (\ref{v_eqn}) with $\partial_x$ leaving
\beq
\frac{D\partial_x v}{Dt} + \left[\Omega_0(2-q) + \partial_x v\right](\partial_x u + \partial_y v) = 0.
\eeq
Using (\ref{anelastic_eqn}) in the above and combining the result with (\ref{h_eqn})
finally results in (\ref{DSWE_1}).
%\beq
%\frac{D\Xi}{Dt} = 0, \qquad
%\Xi \equiv \frac{\Omega_0(2-q) + \partial_x v}{h^{\frac{\gamma+1}{\gamma-1}}}.
%\label{Taylor_Proudmann}
%\eeq
The quantity $\Xi$ is the potential vorticity and (\ref{DSWE_1})
is the analogous Taylor-Proudman theorem for this shallow water system
(cf. Pedlosky, 1987, Cho et al., 2007).  We may further isolate the radial velocity in
terms of $h$.  We do this by (i) multiplying (\ref{h_eqn}) by
 $h\Omega_0^2/2$, (ii) operating on the resulting equation
with the partial derivative operation $\partial_x$, (iii)
subtracting the resulting equation from (\ref{v_eqn}), (iv)
reordering the result and using (\ref{semi_geostrophy}) results
in the diagnostic equations for $u$ and $v$ found in
(\ref{DSWE_2}-\ref{DSWE_4}).
%\beqa
%& & {\bf {\cal P}} u = \nonumber \\
%& & \ \
%(\partial_x {\cal V} - q\Omega_0 - 2\Omega_0)\cdot\partial_y Q_h
%+2\partial_x\left(\frac{\gamma-1}{\gamma+1}Q_h \partial_y {\cal V}\right),
%\eeqa
%in which
%\beqa
%v &=& {\cal V}(h) \equiv \frac{\Omega_0}{4}\partial_x h^2, \qquad
%Q_h \equiv \sfrac{1}{2}\Omega_0^2 h^2,
%\eeqa
%and where the operator ${\cal P}$ is defined as
%\beqa
%{\bf {\cal P}} &\equiv& \Omega_0^22(2-q) - 2\Omega_0 {\cal V}\cdot \partial_x
%-2\partial_x\left(\frac{\gamma-1}{\gamma+1}Q_h \partial_x \right).
%\eeqa
%Note that the semigeostrophic
%balance term relating $h$ and $v$
%is $2\Omega_0 v = \partial_x Q_h$.
%We have recovered the asymptotic
%equations presented in the Introduction.
\section{Constant uniform potential vorticity solutions}
We consider the dynamics emerging from a situation where
$\Xi$ is uniform and constant.  We imagine that we disturb
a uniform equilibrium $h = h_0$.  Thus this means
that the uniform potential vorticity state is given by
\beq
\Xi_0 = \Omega_0\frac{2-q}{h_0^{\frac{\gamma+1}{\gamma-1}}}.
\eeq
Through the use of (\ref{semi_geostrophy}) we find
that the equation to be solved is
\beq
\Omega_0(2-q) + \Omega_0 \partial_x^2 \left(\sfrac{1}{4} h^2\right)
-\Xi_0h^{\frac{\gamma+1}{\gamma-1}} = 0,
\label{PV_eqn}
\eeq
subject to specified boundary conditions.  Given that we
will be considering disturbances around the uniform
state we define a new quantity $\Theta$ as
\[
\theta = \frac{h-h_0}{h_0}
\]
Thus, the potential vorticity equation (\ref{PV_eqn}) is written
more simply as
\beq
\sfrac{1}{2}h_0^2\partial_x^2 \left(1+\theta\right)^2 +
2(2-q)\left[1-(1+\theta)^{\frac{\gamma+1}{\gamma-1}}\right] = 0.
\label{exact_uniform_PV}
\eeq

\par
We shall consider here only two possible
boundary conditions.
The first is that the normal velocity is zero at the position $x=0$ and that
the pressure remain fixed as $x\rightarrow\infty$ which translates
to requiring $h\rightarrow h_0$.  We shall refer to this scenario
as the ``semi-infinite domain".
Referring to (\ref{v_eqn}) the condition on the normal
velocity is the same as requiring
\beq
(\partial_t - q\Omega_0x\partial_y)v + v\partial_y v +\sfrac{1}{2}\partial_y\Omega_0^2 (1+\theta)^2 = 0,
\label{zero_u_bc}
\eeq
at $x=0$ and where $v = \Omega_0\sfrac{1}{4}\partial_x(1+\theta)^2$.
The second set of boundary condition is that we require the normal
velocities be zero at the positions $x=\pm\Delta$.  In other words, we
require that (\ref{zero_u_bc}) be satisfied at those boundaries.  We refer
to this as the ``channel flow".
%\subsection{Linear Theory}
\par
We consider linear disturbances for $\theta$.  The governing
equation for the disturbance PV, $\Xi'$, becomes
\beq
\Xi' \equiv h_0^2\partial_x^2\theta -
2(2-q)\frac{\gamma+1}{\gamma-1}\theta = 0.
\label{Xi_prime}
\eeq
The normal velocity condition at the boundaries becomes
\beq
(\partial_t - q\Omega_0x\partial_y)\partial_x\theta
+ 2\Omega_0\partial_y\theta = -2(2-q)u = 0.
\label{linear_bc_condition}
\eeq
\par
{\em Semi-infinite domain.} The solution appropriate to the
semiinfinite domain is
\beq
\Theta = -\frac{A(y,t)}{\kappa}e^{-\kappa x}
,\qquad \kappa^2 = \frac{2(2-q)}{h_0^2}\frac{\gamma+1}{\gamma-1},
\label{lin_semiinfinite_solution}
\eeq
where we select the $\kappa > 0$ solution.  The amplitude time evolution of the amplitude $A$ is
obtained from the velocity boundary condition (\ref{linear_bc_condition}),
\beq
\partial_t A + 2\Omega_0\kappa^{-1}\partial_y A = 0.
\eeq
 {Making the normal mode ansatz of the form $A(y,t) \sim
\exp{i\alpha(y-ct)}$ where $\alpha$ is the streamwise wavenumber, }we find that the
wavespeed is given by
\beq
c=-2\Omega_0/\kappa. \label{semi-infinite_wavespeed}
\eeq
Since $c<0$, it follows
that these are wave patterns which propagate
in the negative $y$ direction, { that is to say,
against the mean flow.}
\par
{\em Channel flow}.  Under channel flow conditions
the normal-mode solution of $\theta$ is
\beq
\theta = \Bigl[A\cosh \kappa(x+\Delta)
- B\cosh\kappa(x-\Delta)\Bigr]e^{i\alpha(y-ct)}.
\label{channel_flow_Theta_lin}
\eeq
Enforcing the linear boundary conditions
(\ref{linear_bc_condition})
at the points $x=\pm\Delta$ using the solution
(\ref{channel_flow_Theta_lin}) we find
\beq
-(c+q\Omega_0\Delta)A\kappa \sinh{2\kappa\Delta}
+ 2\Omega_0(A \cosh{2\kappa\Delta} - B) = 0,
\eeq
at $x=-\Delta$ and
\beq
-(c-q\Omega_0\Delta)B\kappa \sinh{2\kappa\Delta}
+ 2\Omega_0(A - B\cosh{2\kappa\Delta}) = 0,
\eeq
at $x=\Delta$.
These two simultaneous equations are satisfied non-trivially if
the wave speed $c$ satisfies,
\beq
c^2 = q^2\Omega_0^2\Delta^2 + \frac{4\Omega_0^2}{\kappa^2}
\Bigl(1-q\kappa\Delta\coth 2\kappa\Delta\Bigr).
\label{sri_wavespeed}
\eeq

\begin{figure}
\begin{center}
\leavevmode \epsfysize=7.5cm
\epsfbox{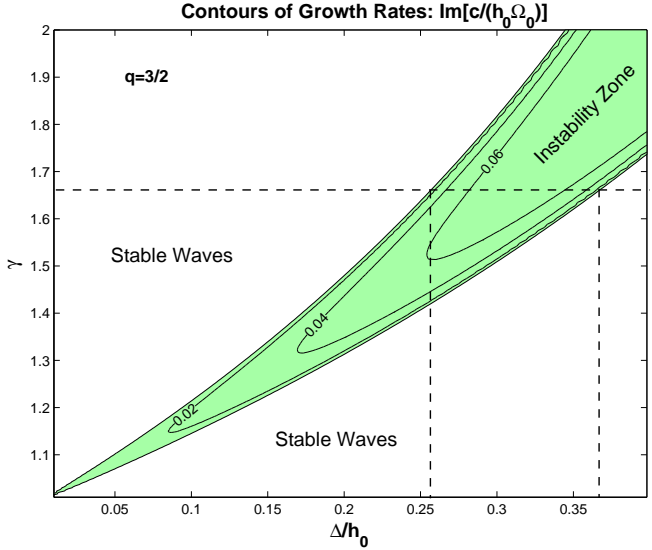}
\end{center}
\caption{Contours of growth rates of the SRI as a function of
$\gamma$ and $\Delta/h_0$.  The growth rates are scaled by $\Omega_0 h_0$.
The behavior for $q=3/2$ is shown.  The shaded region corresponds to the
instability band.  Horizontal line designating $\gamma = 5/3$ is also shown.}
\label{contour_ass}
\end{figure}

The wavespeeds approach the
local mean flow speed of the respective boundaries as the separation
between the walls becomes large, i.e.
\beq
\lim_{\Delta\rightarrow\infty} c^2 = \Omega_0^2\Delta^2 q^2.
\eeq
On the other hand, when the wall separation becomes small
the wavespeeds limit to
 \beq
\lim_{\Delta\rightarrow 0} c^2 = h_0^2\Omega_0^2 \frac{\gamma-1}{\gamma+1}.
\eeq
Examining the wavespeed prescription (\ref{sri_wavespeed}) in the
case where there is no shear, i.e. $q=0$, shows that there
is no possibility of instability.
However, we do see that there seems to exist a band of instability
in the gap width $\Delta$ for values of $q>0$.  In Figure \ref{contour_ass} we plot
the region of instability and growth rate as a function of a variety
of parameters.  We see that range in $\Delta$ for which instability is
predicted is less than $h_0$: for instance, for $\gamma = 5/3$ instability
is expected approximately only when $0.26 h_0  < \Delta < 0.37 h_0$.   The range widens
as $\gamma$ increases but is still less than $h_0$ for physically
reasonable values of $\gamma$.  Note that the thickness
of the channel is measured by $2\Delta$.

\section{Normal-modes of an infinite domain with a single PV defect}
In this section we consider the response of an infinite
domain separated by a single defect in the potential vorticity.
For simplicity we again treat $h_0$ as constant.
Therefore, the defect appears in this case
as a jump in the basic
state shear, i.e.
\beq
q = \left\{
\begin{array}{ll}
q_+, & \ \ x>0, \\
q_-, & \ \ x<0,
\end{array}
\right. \ ,  \qquad
V_0 = \left\{
\begin{array}{ll}
-\Omega_0 q_+ x, & \ \ x>0, \\
-\Omega_0 q_- x, & \ \ x<0,
\end{array}\right. .
\eeq
Where $V_0$ is the steady state uniform shear.  This formulation
ensures that the background velocity is everywhere continuous.
Normal mode disturbances can be developed for this configuration much
in the manner developed in the previous section.  This means
separately solving
(\ref{Xi_prime}) on either side of the defect and then
matching the solutions across the interface.  \footnote{
Another way to look at this would be to argue that if the
perturbed height of the disk were to change discontinuously
then it would correspond to a delta-function horizontal
pressure gradient.}
Thus, for
$x>0$, disturbances which decay as $x\rightarrow \infty$
are
\beq
\theta = A_+ e^{i\alpha(y-ct)}e^{-\kappa_+ x}, \qquad \kappa_+^2 = \frac{2(2-q_+)}{h_0^2}\cdot
\frac{\gamma + 1}{\gamma-1},
\eeq
while for
$x<0$, disturbances decaying as $x\rightarrow -\infty$
are
\beq
\theta = A_- e^{i\alpha(y-ct)}e^{\kappa_{_{-}} x}, \qquad \kappa_-^2 = \frac{2(2-q_-)}{h_0^2}\cdot
\frac{\gamma + 1}{\gamma-1},
\eeq
while in both cases the positive root for $\kappa_\pm$ is chosen.
Since $\theta$ represents the perturbed height of the disk
we require that it be continuous across the interface - this
results in $A_- = A_+ = A$.  Because we are investigating
normal mode disturbances, it also makes physical sense
to match the normal velocities across the interface.
There are a number of ways to do this but the mathematically
simplest is to consider the linearized version of
(\ref{y_momentum_eqn}),
\beq
(\partial_t - q\Omega_0 x\partial_y)v + \Omega_0(2-q) u' = - h_0\partial_y \theta.
\label{linearized_u}
\eeq
Since the linearized expression of the
geostrophic balance
relationship is $v' = \sfrac{1}{2}\partial_x \theta$,
in order for the radial velocity $u'$ to be continuous across the
defect we must have that
\beq
\Bigl[u'\Bigr]^{x \rightarrow 0^+}_{x\rightarrow 0^-} =
\left[
\frac{i\alpha(-c-q\Omega_0 x)\partial_x \theta + 2\Omega_0i\alpha\theta}{2-q}
\right]^{x \rightarrow 0^+}_{x\rightarrow 0^-} = 0.
\label{jump_condition}
\eeq
Given the solution for $\theta$ on either side of $x=0$, in order
for this to be satisfied is if the wavespeed $c$ satisfies the
relationship
\beq
c = -2\Omega_0\frac{q_+-q_-}{\kappa_+(2-q_-) + \kappa_-(2-q_+)}. \label{defect_wavespeed}
\eeq
The wavespeed represents the propagation of a Rossby edgewave
in the sense presented in Hoskins et al. (1985).  So long as $q_\pm <2$
the edgewave propagates in a negative sense with
respect to the coordinate $y$ if $q_+ > q_-$.  Stated in another
way, when this condition is met the Rossby wave travels against
the mean flow.  \par
We note a significant
mathematical limit of the wavespeed prescription (\ref{defect_wavespeed}):
when $q_- \rightarrow -\infty$,
while keeping $q_+$ finite, we find that
\beq
c \rightarrow -2\Omega_0/\kappa_+.
\eeq
By identifying $q_+$ with $q$ it follows from
the definition of $\kappa$ (\ref{lin_semiinfinite_solution})
that $\kappa_+ = \kappa$.  Hence,
we have recovered the wavespeed form for the semi-infinite domain,
i.e. (\ref{semi-infinite_wavespeed}) in which the edgewave propagates against the local mean flow.
This view allows us to interpret the presence
of an inner wall boundary condition to be
like a PV-defect whose PV on the one side of
the defect (e.g. $x<0$)
greatly exceeds in magnitude the PV of the other side of
the defect.  \par
This implication can also be understood by
studying (\ref{jump_condition}): when, say, $-q_- \gg 1$ then
it means that the normal velocity approaching the defect from $0^+$
should be correspondingly small.  In the infinite
limit it is equivalent to requiring $u'(x\rightarrow 0^+) = 0$.

\section{Linear, non-uniform potential vorticity solutions}
We consider here linearized solutions to the
potential vorticity $\Xi'$ when it is not
in general zero.  This means solving
\beq
(\partial_t - q\Omega_0 \partial_y)\Xi' = 0
\eeq
given the definition of $\Xi'$ to be found
in (\ref{Xi_prime}), this may be
rewritten instead as
\beq
(\partial_t - q\Omega_0 x \partial_y)
\left[
2(2-q)\frac{\gamma+1}{\gamma-1} - h_0^2\partial_x^2
\right]\theta = 0.
\eeq
The radial velocity is the solution to
\beqa
& & \left[2(2-q) - \frac{\gamma-1}{\gamma+1}h_0^2\partial_x^2
\right]u = \nonumber \\
& & \ \ \ \ h_0^2\Omega_0\left[-2(q + 2) +
\frac{\gamma-1}{\gamma+1}
h_0^2\partial_x^2\right]\sfrac{1}{2}\partial_y \theta.
\label{lin_u_relationship}
\eeqa
We shall consider here wave solutions on
the classical sheared periodic domain
as introduced by Goldreich \& Lynden-Bell (1965).
To this end we define a new coordinate system in which
the new azimuthal coordinate moves with the local shear
\beq
X = x, \qquad Y = y + q\Omega_0x t, \qquad T = t,
\eeq
so that all derivative operations now translate
to
\beq
\partial_t \rightarrow \partial_T,\qquad
\partial_x \rightarrow \partial_X + q\Omega_0 T \partial_y, \qquad
\partial_y \rightarrow \partial_Y.
\eeq
Then in the usual way we find the evolution equation
for $\Xi'$ can now be written as
\beq
\partial_T\left[h_0^2 (\partial_X + q\Omega_0 T\partial_Y)^2
+ 2(2-q)\frac{\gamma+1}{\gamma-1}\right]\theta = 0.
\eeq
In this frame we assume Fourier solutions of the form
\beq
\theta = \theta_{k\ell}(T)e^{ikX + i\ell Y},
\eeq
where $k$ and $\ell$ are wavevectors.  The solution
to the wave amplitude $\theta_{k\ell}(T)$ is given
simply by
\beq
\theta_{k\ell} = \hat\theta_{k\ell}
\frac{h_0^2k^2 + 2(2-q)\frac{\gamma+1}{\gamma-1}}
{h_0^2(k+q\Omega_0\ell T)^2 + 2(2-q)\frac{\gamma+1}{\gamma-1}},
\eeq
where $\hat\theta_{k\ell}$ is the wave amplitude at time
$T=0$.  The solution to the aziumuthal velocity
may also be expressed as
\beqa
v &=& v_{k\ell}e^{ikX + i\ell Y}, \nonumber \\
v_{k\ell} &=& -i\sfrac{1}{4}\Omega_0 h_0 (k+q\Omega_0\ell T)
\theta_{k\ell}.
\eeqa
Similary, representing the solution for
$u$ as
\[
u_{k\ell}(T) e^{ikX + i\ell Y},
\] we have from
(\ref{lin_u_relationship})
\beq
u_{k\ell} =
-i\ell\sfrac{1}{2}h_0^2\Omega_0
\frac{2(2+q) + \frac{\gamma-1}{\gamma+1}h_0^2
(k+q\Omega_0\ell T)^2}
{2(2-q) + \frac{\gamma-1}{\gamma+1}h_0^2
(k+q\Omega_0\ell T)^2}\theta_{k\ell}.
\eeq

Long time evolution of the solution to $\theta_{k\ell}$
and $u_{k\ell}$
shows decay as $1/T^2$ while $v_{k\ell}$ decays as $1/T$.
If one restricts attention
to solutions where $T>0$,
leading waves, i.e. those in which $k\ell < 0$, will
show a maximum in the amplitude of $\theta_{k\ell}$
when $T_{{\rm max}} = k/(q\Omega_0\ell)$.
The ratio of the amplitudes of these trailing waves
to their initial amplitude is
\[
\frac{\theta_{k\ell}(T_{{\rm max}})}{\theta_{k\ell}(0)}
= 1 + \frac{h_0^2 k^2}{2(2-q)}\cdot \frac{\gamma-1}{\gamma+1},
\]
showing that amplification is greater as $k$ increases.
Furthermore, at $T_{{\rm max}}$ we see that $v_{k\ell} = 0$.
\par
On the other hand, trailing waves, i.e.
those in which $k\ell > 0$, show monotonic decay for
all $T>0$.
Examples of this sort of behavior is well-known
for the evolution of two dimensional vorticity
perturbations (e.g. Chagelishvilli et al. 2003,
Umurhan \& Regev, 2004) and three-dimensional
disturbances (Knobloch, 1984, Korycansky, 1992,
Sternberg, 2005, Balbus \& Hawley, 2006).
\section{Discussion and Summary}
In 2D inviscid, incompressible vortex flow in a
background linear shear profile, the vorticity is
a quantity which is conserved in a Lagrangian
sense.  In terms of the non-dimensional
quantities we have come to use in this study,
this means to say that that
\beq
\left[\partial_t + u\partial_x + (v-q\Omega_0 x)\partial_y\right]\omega = 0,
\eeq
in which the vorticity $\omega$ may be expressed in
terms of a stream-function $\psi$ according to
$
\omega = (\partial_x^2 + \partial_y^2)\psi.
$
The radial and azimuthal velocities are then
given by
$
u = \partial_y\psi, \ v_y = - \partial_x\psi.
$
By construction the $u$ and $v_y$ velocities
satisfy the criterion of incompressibility.
\par
By comparison, the asymptotic system we have developed here,
representing the varicose disturbances of an
annular section of a cold rotationally balanced
accretion disk containing a homentropic gas, has a similar form.
In other words it means to say that the corresponding
Lagrangian conserved quantity is the potential vorticity
$\Xi$, i.e.
\beq
\left [\partial_t + u\partial_x + (v-q\Omega_0 x)\partial_y\right]\Xi = 0.
\eeq
The potential vorticity is related
to the disk height $h$ according to
\beq
\Xi =\frac{\Omega_0(2-q) + \partial_x {\cal V}(h)}{h^{{\frac{\gamma+1}{\gamma-1}}}},
\label{PV_again}
\eeq
where $\gamma$ is
the polytropic index characterizing the homentropic layer under consideration.  The radial
and vertical velocities are given by
\beq
v={\cal V}(h),\qquad u = {\cal P}^{-1}{\cal F}(h),
\eeq
in which the operators appearing above are given in (2-4).
The disk height $h$ is, in this case, analogous to the
streamfunction in classical 2D inviscid incompressible
theory.
Unlike the latter,
the horizontal velocities
reflect the dynamics of a quasi-anelastic relationship
of the flow quantities in 3D, i.e.
\beq
\partial_x u + \partial_y v_y +\frac{\gamma+1}{\gamma-1}\Omega_z = 0,
\label{anelastic_again}
\eeq
where in general the vertical velocity
is given as $w = \Omega_z z$.  The relationships
(\ref{PV_again}-\ref{anelastic_again}) including the
definition of the operators (2-4) are self-contained.\par
Note that the quantity $\Omega_z$ does not play a direct
role in the evolution of the potential vorticity.  Its meaning
can be understood by noting
that the expressions establishing
the relationships between the horizontal velocities and
the disk height $h$ are consistent with the material
derivative of the disk height, that is, its role may be
understood in terms of the relationship,
\beq
\left [\partial_t + u\partial_x + (v-q\Omega_0 x)\partial_y\right] h
= w(z=h) = h\Omega_z.
\eeq
\subsection{Relationship to other approximations}
In Table 1 we present a comparison of the various approximations
used to model small sections of accretion disks (i.e. conditions
in which $\epsilon \ll 1$).
The four
approximations presented are ones used
in the literature and we describe them briefly below.  In all
of the following standard approximations the vertical extent of
the disk $H$ is assumed to be smaller than the radial
scales of the disk by order $\epsilon$ - that is, $H \sim
\order{\epsilon R_0}$.  Additionally, and within the moving
frame assumed throughout these considerations, a Mach number
is defined by, $\delta \equiv {\cal U}/c_s$, where ${\cal U}$
is the largest velocity scale encountered in the frame.
\begin{description}
\item[$\bullet$] \emph {SSB.} \ \  The small shearing box equations are the classical
ones used to investigate the dynamics on very small scale sections
of disks.  Originally proposed by Goldreich \& Lynden-Bell (1965)
they have been the basis of numerous studies pertaining to the
question of hydrodynamic disk turbulence (Lesur \& Longaretti, 2005)
and nature and maintanence of 3D disk vortices (Lithwick, 2007 B).
A scaling analysis (Umurhan \& Regev, 2004) shows
that these disturbances are assumed to have Mach numbers smaller
than the aspect ratio of disks (i.e. $\delta \ll \epsilon$).  The
corresponding length scales of disturbances are smaller than
the vertical extent of the disk  as well but the time scales
of all disturbances are equal to the local disk rotation times (i.e. $\sim
1/\bar\Omega_0$).
\item[$\bullet$] {\emph{LSB.}} \ \ The so-called large shearing box equations are
derived from a scaling analysis similar to those leading to the SSB
(Umurhan \& Regev, 2004) except that the velocities are scaled
to have Mach numbers which are order 1.  The corresponding length
scales of disturbances compare to the vertical scale of the disk.
As with the SSB, the time scales of disturbances equal the
local disk rotation times.
The resulting equations are the compressible version of the equations
of the SSB which also allow for the effects of gravity waves as
there is now also an entropy equation which is evolved as well.
\item[$\bullet$] {\emph{Elongated Vortex.}}
\ \ Assuming a flow Mach number which is small,
an analysis of a Keplerian disk in which one considers
disturbances whose azimuthal and vertical scales compare
to the local disk height while the radial length scales
are smaller by an order $\delta$ lead to the equations
describing elongated vortices (Barranco, et al., 2000).
Under these circumstances, while the subsonic
horizontal and vertical velocities of disturbances
scale by the Mach number of the soundspeed (i.e. $\delta c_s$)
the scale of the
radial velocity disturbances are smaller than these by another
order of the Mach number.  Consequently the time
scale of all disturbances are an order $1/\delta$
longer than the local disk rotation time.
\item[$\bullet$] {\emph{Anelastic Model.}}
Barranco \& Marcus (2005) present a scaling argument
leading to a set equations which we refer
to here as the Anelastic Model.  Flow quantities
are formally assumed to have small Mach numbers
(i.e. $\delta \ll 1$).  Azimuthal and radial
disturbance scales are assumed to be an order
$\delta$ smaller than the vertical scale of the disk $H$
while the vertical disturbances scales are of
the order $H$.  The subsonic azimuthal and radial velocity
disturbances scale by the Mach number factor of the
soundspeed while the vertical velocities are a factor
of $\delta$ even smaller than these. Consequently, the time
scales of all these vortical disturbances
is of the order of the disk rotation time.  By assuming
that the dynamical pressure and densities scale as
they do one arrives at a circumstance in which
sound waves are filtered from the equation set
resulting in
equations of motion describing an anelastic fluid.
\end{description}
The scalings describing the dynamical response of
the fluid pressure and density disturbances are
also presented in the table for the purpose of
comparison.  The point to take away from
this is that the pressure and density fluctuations
are small by order $\delta^2$ their respective
steady state values in the equations describing elongated
vortices and the Anelastic Model as compared to
the corresponding pressure
and density disturbances considered in this work and for
the LSB.  The dynamical pressure fluctuations also
scale as a factor of $\delta$ in the SSB.  The small
scalings of the thermodynamic quantities affect
the filtering out of soundwaves.
\par
The scaling analysis we have employed here to arrive
at this shallow water limit have some features
similar to the arguments leading to the the Elongated
Vortex equations of Barranco et al. (2000).  Physically
it means that
there exists radial geostrophic balance of dynamical quantities
- a condition which has been visited and explored in prior studies
(Balmforth et al., 1992, Balmforth \& Spiegel, 1996).
The important feature that the approximations made in this
study shares with the arguments leading to the Elongated Vortex
equations is that the dynamical timescales of the disturbances
are long compared to the local disk rotation time.  This
allows for an analysis which can focus on the long term
properties of vortices under these circumstances free of the
effect of soundwaves.  Soundwaves and other short time
fluctuations in this study
have been filtered out by the fact that we are
considering time scales which are longer than
what typifies these wave phenomena.
  It is this long timescale view together with the assumption made
here that
the radial and vertical velocity fluctuations are smaller
than the azimuthal ones that ultimately lead to the shallow water
theory limit arrived at here.
\begin{table}
\caption{Comparison of the various scalings employed
that result in some of the approximations used in the literature
to study small sections of accretion disks.}
\centering
\begin{tabular}{l c c c c c}
\hline\hline
 & & & Elongated & Anelastic & This  \\
Quantity & SSB & LSB & Vortices & Model & Work \\
\hline
\hline
\\
$\epsilon \equiv c_s/v_k$ & $ \ll 1$ & $ \ll 1$ &
$ \ll 1$
& $ \ll 1$ & $ \ll 1$ \\
\\
$\delta$: Mach \# & $ \ll \epsilon$ &
$\sim 1$ & $\epsilon \ll \delta \ll 1$ & $\epsilon \ll \delta \ll 1$ &
$\sim 1$ \\
\hline
\\
azimuthal & & & & & \\
scales & $\delta R_0$ & $\epsilon R_0$
& $\epsilon R_0$ & $\delta\epsilon R_0$ & $R_0$ \\
\\
vertical & & & & & \\
scales & $\delta R_0$ & $\epsilon R_0$ & $\epsilon R_0$ &
$\delta \epsilon R_0$ & $\epsilon R_0$ \\
\\
radial & & & & & \\
scales & $\delta R_0$ & $\epsilon R_0$ & $\delta\epsilon R_0$ &
$\delta\epsilon R_0$ & $\epsilon R_0$ \\
\\
azimuthal & & & & & \\
speeds & $\delta c_s$ & $c_s$ & $\delta c_s$ &
$\delta c_s$ & $c_s$ \\
\\
vertical & & & & & \\
speeds & $\delta c_s$ & $c_s$ & $\delta c_s$ &
$\delta c_s$ & $\epsilon c_s$ \\
\\
radial & & & & & \\
speeds & $\delta c_s$ & $c_s$ & $\delta^2 c_s$ &
$\delta c_s$ & $\epsilon c_s$ \\
\\
temporal & & & & & \\
scales & $1/\bar\Omega_0$ & $1/\bar\Omega_0$ &
$1/(\delta\bar\Omega_0)$ &
$1/\bar\Omega_0$ & $1/(\epsilon\bar\Omega_0)$ \\
\\
dynamical & & & & & \\
pressure & $\delta\bar\rho c_s^2$ & $\bar\rho c_s^2$ &
$\delta^2\bar\rho c_s^2$ &
$\delta^2\bar\rho c_s^2$ & $\bar\rho c_s^2$ \\
\\
dynamical & & & & & \\
density & $\bar\rho$ & $\bar\rho$ &
$\delta^2\bar\rho $ &
$\delta^2\bar\rho $ & $\bar\rho $ \\
\hline
\end{tabular}
\end{table}
\subsection{Walls, defects, edgewaves and instability}
 {In an examination of single PV defect disturbances of
an infinite domain we saw that by letting the PV on
one side of the defect tend to infinity is
mathematically equivalent to replacing the defect
with an impenetrable wall.  Indeed, the limiting form
of the Rossby edgewave wavespeed (\ref{defect_wavespeed}) in the limit where
the PV on one side becomes infinite recovers the expression for
the wavespeed
on the semi-infinite domain (\ref{semi-infinite_wavespeed}).  It would
seem then, at least within the context of this theory, that this
equivalency may be a useful conceptual tool.  For instance, the use
of impenetrable boundary conditions is widely considered to be
artificial in most cases other than for terrestrial laboratory flows
in a channel.
This consensus is especially true in the study of astrophysical disks.
Of course as disks no not have delineated solid boundaries, this
view is physically justified.  However, it is conceivable that
disks may have regions in which the radially varying potential
vorticity undergoes rapid changes.  One obvious candidate
region would be the star-disk boundary.  If the star rotates
significantly sub-Keplerian then the PV in the boundary layer region
is likely to be significantly different
than the PV immediately outside the boundary layer.  The transition
between the two can then be viewed as a defect in the disk's PV
and in the limit where the PV in the boundary layer greatly exceeds the
PV outside, the arguments laid out here could be used
to justify replacing
the defect with a solid boundary for the purposes of analyzing the
response in the extended disk.}
\par
 {With this in mind, we can view the edgewave propagating along the solid
boundary as a limiting form of a Rossby edgewave propagating along
a defect.  It follows that much of the machinery involved
with the principle of counterpropagating Rossby waves (CRW),
originally proposed in Hoskins et al. (1985) in the context
of understanding barotropic/baroclinic instabilities in planetary atmospheres,
should carry over to understanding disks as well.
In Section 3 we presented a normal mode analysis on a finite
domain bounded by two impenetrable walls.  The instability
associated with it is a manifestation of the SRI.
In view of the analysis, we see that the two modes appearing
under those circumstances are Rossby edgewaves propagating along the
boundary walls.  Under the right
conditions (i.e. wall separation) these edgewaves interact with each other
in such a way as to promote an instability in the way outlined and
demonstrated in Molemaker et al. (2001) - much in the same way
that edgewaves are responsible for the emergence of the PPI
(Goldreich et al., 1985, and see below).}
\par
 {Now, given the foregoing
argument, the edgewaves associated with the
boundaries can be viewed as a limiting form of
Rossby edgewaves propagating along a PV defect in which
the PV on one side greatly exceeds the other.  In this way, the left boundary
looks like it has a negative PV gradient while the gradient on
the right boundary looks positive which satisfies the two-signed
PV-gradient instability condition of Charney \& Stern (1962)
which means that
%Such circumstances
%satisfies the necessary Charney-Stern criterion for baroclinic
%instability which states that there is instability if there is a %change of sign
%in the PV-gradient somewhere on an isentropic surface
%(Charney \& Stern, 1962).  It would seem, therefore, that the SRI
%shares mathematical similarities to baroclinic instabilities
%typically encountered in planetary atmosphere studies and
the mechanism of the instability
can be physically rationalized as arising from the interaction of CRWs
(Hoskins et al., 1985, Heifetz et al., 2004).
Such instabilities
ultimately draw their energy from the background shear - without which
the instability vanishes as is the case here.
An examination of the SRI in which the channel boundaries are
replaced with either PV defects or strongly varying smooth PV profiles
is the subject of a future study.
}\par
 {A real disk probably is rotationally supported all the way through, which
would mean that the only way that one could realistically argue
for significant changes in its PV profile would be if the steady-state height of the
disk changes as a function of the radial coordinate.
Such localized annuli will probably introduce some departures
from Keplerian flow as well.  The possibility of this scenario
has been proposed before by Lovelace et al. (1999) and Li et al. (2000) in the context of
two-dimensional disks characterized by strong radial entropy gradients
where they show that Rossby waves turn unstable under suitable circumstances.
It is in the opinion of this author that the viability of this process
shows promise and ought to be further considered.}

\subsection{Dynamical results in comparison to other studies}
The linear instability discovered by Papaloizou \& Pringle (1984,1985)
was analytically demonstrated in an analysis of homentropic gas in slender
accretion tori (i.e. not exactly rotationally supported) constrained by
pressure conditions
on the boundaries of the domain.  This means to say that the vertical and radial
scales are much smaller than the azimuthal scales
and the flow is contained in a narrow annular channel.  Furthermore,
the modal disturbances have very nearly uniform potential vorticity throughout
and have the character of being essentially vertically hydrostatic
as illucidated by Goldreich et al (1986).
This instability occurs for sinuous modes (i.e. having even symmetry with respect
to the disk midplane) while no such instability is observed for varicose
modes (i.e. having odd with respect to the disk midplane).
The instability has been physically rationalized
by understanding that the pressure
boundary conditions applied at the inner and outer walls of the
undulating tori essentially set up a situation in which
edgewaves counterpropagate with respect to each other and interact
across a region which is evanescent - leading ultimately to instability
(Goldreich et al, 1986, Blaes \& Glatzel, 1986).  In view of this
mechanistic understanding, the PPI should be considered the disk analog
of Eady edgewaves encountered frequently in terrestrial atmospheric
studies (e.g. Davies \& Bishop, 1994).
\par
The dynamics considered here are for rotationally supported
disks ($q=3/2$) which differ mathematically from tori ($q\neq 3/2$) which
have significant radial pressure support at the lowest order of the dynamics.
The instability of uniform potential vorticity modes
discussed in Section 3 and which is an incarnation of the
strato-rotational instability (SRI, Dubrulle, 2004) is not
the same as the Papaloizou-Pringle instability (PPI).  The main
differences are (i) the symmetry of the modes discussed
in this work are
varicose while they are sinuous for the PPI and (ii)
the PPI arises in part from the pressure boundary conditions at
the inner and outer boundaries
while the instability here arises from the no normal flow condition
at the inner and outer boundaries.
Both instabilities, however, share (i) the characteristic as
arising from the interaction of counterpropagating edgewaves
of some sort and are explained in terms of the Hayashi-Young
criterion for wave instability in shear flow
(Hayashi
\& Young, 1987) and, (ii) the character of belonging to a general
class of barotropic/baroclinic instabilities related to the dynamics
of potential vorticity disturbances (Bretherton, 1966, Hoskins et al., 1985, Davies \& Bishop, 1994).
%It was also shown in Papaloizou \& Pringle (1985)
%that varicose modes of the accretion tori are stable.
%and this is probably the case because the nature of
%disturbances, and in particular, the effect of
%the pressure boundary conditions, drastically changes
%because the effective radial gravity vanishes in the
%limit of
%Furthermore, a cursory review of the instability condition of the PPI in the narrow
%annular limit (cf. Goldreich
%et al., 1986) shows it tends to zero as $q\rightarrow 3/2$.
\subsection{Future directions}
A natural test and evaluation of the shallow water
theory developed in this study will be to study the nonlinear
solutions of non-uniform potential vorticity disturbances.
At the outset such an investigation will surely require
numerical computational simulations.  The question of what
boundary conditions to use will be a sensitive one since,
for instance, certain conditions
may lead to strong linear
instabilities and the relevance of them to real disks
are questionable (although certainly not yet settled).
For example, by requiring no-normal flow boundary conditions
can excite disturbances which include
uniform potential vorticity perturbations which have
been shown to be the analog of the SRI.   {However, in light of the arguments presented here,
it is possible that
such boundary conditions and their resulting dynamical response could
be qualitatively representative of what goes on in bonafide real disk systems.}
\par
It should also be considered that
the shallow water theory developed here (by its construction)
is unable to follow
the dynamical response of disturbances in which there is a vertical
variation in the entropy on account of the assumptions made
about the thermodynamical response of the fluid made at the
outset (i.e. the assumption of a polytropic gas).  Vertical
variations of this sort may be playing a critical role in
the development of the dynamics reported in Barranco \& Marcus (2005)
since it was observed in that work that vertically propagating
disturbances contribute to the complex vortex dynamics seen
to manifest in their simulations ( {see also the discussion on
the stability of
tall columnar vortices in Dritschel \& Juarez, 1996}).  As one of the long term goals
we have in mind for developing a shallow water formalism
of non-axisymmetric disturbances is to better understand
the emergence of such interesting dynamics,
it will be worthwhile to extend the theory developed here
in order to accommodate such circumstances by allowing for vertical
variations of some type beyond the fluctuations of the disk
height we follow here.  One way would
be to develop a multi-stacked shallow water theory in a manner
similar to what has been developed in the study of MHD
disturbances in the solar tachocline (Gilman, 2000, Cho et al., 2007).\par
We would like to make one final observation pointed out by the referee
of this article.  With respect to the problem of
stellar tachoclines Zaqarashvili et al. (2007) discuss
how the inclusion of a background azimuthal magnetic field can split
an ordinary Rossby wave into two different parallel modes.  By analogy
a similar phenomenon is likely to occur in magnetized disks
(see also Matsumoto \& Tajima, 1995).  A strategy like the one
employed in this work may be used to further detail the scope
and shape of the modification of Rossby waves in such circumstances.
\section{Acknowledgements}
The author offers thanks to the anonymous referee who offered
numerous suggestions to improve the quality of the presentation of this work.
The author also thanks James Cho for a number of enlightening
discussions.  This research was supported in part by BSF grant 2004087
and ISF grant 1084/06.

\end{document}